\documentclass[12pt]{iopart}
\usepackage{iopams}
\usepackage{graphicx}
\expandafter\let\csname equation*\endcsname\relax
\expandafter\let\csname endequation*\endcsname\relax
\usepackage{amsmath}
\usepackage{amssymb}

\begin{document}

\title[The distribution of first hitting times of RWs on ER networks]
{The distribution of first hitting times of random walks on 
Erd\H{o}s-R\'enyi networks
}

\author{Ido Tishby, Ofer Biham and Eytan Katzav}
\address{Racah Institute of Physics, 
The Hebrew University, Jerusalem 91904, Israel.}
\eads{\mailto{ido.tishby@mail.huji.ac.il}, \mailto{biham@phys.huji.ac.il}, 
\mailto{eytan.katzav@mail.huji.ac.il}}

\begin{abstract}
Analytical results for   
the distribution of first hitting times of random walks on 
Erd\H{o}s-R\'enyi networks are presented.
Starting from a random initial node,
a random walker hops between adjacent nodes
until it hits a node which it has already visited before.
At this point, the path terminates.
The path length, namely
the number of steps, $d$, pursued 
by the random walker from the initial node
up to its termination is called the first
hitting time
or the first intersection length.
Using recursion equations, 
we obtain 
analytical results for the tail distribution of the path lengths, 
$P(d > \ell)$.
The results are found to be in excellent agreement with numerical
simulations.
It 
is found
that the distribution
$P(d > \ell)$
follows a product of an exponential distribution 
and a Rayleigh distribution.
The
mean, median and standard deviation of this distribution 
are also calculated, 
in terms of the network size and its mean degree.
The termination of an RW path may take place
either by backtracking 
to the previous node
or by
retracing of its path, namely stepping into a node which has 
been visited two or more time steps earlier.
We obtain analytical results for the probabilities,
$p_b$ and $p_r$, that the cause of termination will
be backtracking or retracing, respectively.
It is shown that in dilute networks the dominant termination
scenario is backtracking while in dense networks
most paths terminate by retracing.
We also obtain expressions for the conditional distributions 
$P(d=\ell | b)$ 
and
$P(d=\ell | r)$,
for those paths which are terminated by
backtracking or by retracing, respectively.
These results provide useful insight into the general problem of 
survival analysis and the statistics of mortality rates when
two or more termination scenarios coexist.

\end{abstract}

\pacs{05.40.Fb, 64.60.aq, 89.75.Da}


\vspace{2pc}
\noindent{\it Keywords}: 
Random network, 
Erd\H{o}s-R\'enyi network,
degree distribution,  
random walk, 
self-avoiding walk, 
first hitting time,
first intersection length.

\submitto{\JPA (\today)}
\date
\maketitle

\section{Introduction}

Random walk (RW) models 
\cite{Spitzer1964,Weiss1994}
are useful for the study of a large variety of
stochastic processes 
such as diffusion
\cite{Berg1993,Ibe2013},
polymer structure 
\cite{Fisher1966,Degennes1979},
and random search
\cite{Evans2011,Lopez2012}.
These models were studied extensively 
in different geometries,
including
continuous space
\cite{Lawler2010b}, 
regular lattices 
\cite{Lawler2010a},
fractals 
\cite{ben-Avraham2000}
and 
random networks
\cite{Noh2004}.
In the context of complex networks
\cite{Havlin2010,Newman2010}, 
random walks can be used for either probing the
network structure itself
\cite{Costa2007}
or to model dynamical processes such as the
spreading of rumors, opinions and epidemics
\cite{Pastor-Satorras2001,Barrat2012}.

A RW on a network hops randomly at each time step
to one of the nodes which are adjacent to the current node.
Thus, if the current node is of degree $k$, 
the probability of each one of its neighbors to be 
selected by the RW 
is $1/k$.
Starting from a random initial node,
$x_0$,
the RW generates a path of the form
$x_0 \rightarrow x_1 \rightarrow \dots \rightarrow x_t \rightarrow \dots$,
consisting of the nodes it has visited.
In some of the steps it hops into new nodes which 
have not been visited before.
In other steps it hops into previously visited nodes,
forming loops in its path.
The number of distinct nodes visited up to time $t$
is thus typically smaller than $t$.
The scaling of the mean number of distinct nodes, $s(t)$, visited
by a RW on a random network 
after $t$ steps was recently studied
\cite{Debacco2015}.
It was found that 
for an infinite network
$s(t) \simeq r t$, where 
$r<1$ is a prefactor which depends on the network topology.
These scaling properties resemble those obtained
for RWs on high dimensional lattices.
In particular, it was found that RWs on random networks
are highly effective in exploring the network, retracing their
steps much less frequently than RWs on low dimensional lattices
\cite{Montroll1965}.
In the case of finite networks, another interesting quantity which appears, 
is the mean cover time, namely the average number of steps required
for the RW to visit all the nodes in the network
\cite{Kahn1989}.
Unlike regular lattices, on a complex network of a finite size,
the rates in which the RW visits different nodes are not
identical, but may depend on the degree of the node, its location in
the network and on various correlations between adjacent nodes.
In a random, undirected network which exhibits no correlations, 
such as the  
Erd{\H o}s-R\'enyi (ER)
network,
the rate in which nodes of degree $k$ are visited, is
linearly proportional to $k$.
RWs on random networks also give rise to various first passage
problems
\cite{Redner2001}.
An interesting example is the mean trapping time, 
namely the average number of steps required for a RW in order to reach a specific
node from a random initial node
\cite{Sood2005}.

A special type of random walk, which has been studied 
extensively on regular lattices, 
is the self avoiding walk
(SAW).
This is a random walk which does not 
visit the same node more than once
\cite{Madras1996}.
At each time step, the walker chooses its next move
randomly from the neighbors 
of its present node, excluding nodes
which were already visited
\cite{Fisher1959,Kesten1963,Kesten1964,Hara1993,Clisby2007,Viana2012,Clisby2013}.
The path terminates when the 
SAW
reaches a stalemate situation, 
namely a dead end node
which does not have any yet unvisited neighbors.
The path length, 
$d$,
is given by the number 
of steps made by the RW until it was terminated.
The path length of an SAW on a connected network of size $N$
can take values between $1$ and $N-1$.
The latter case corresponds to 
a Hamiltonian path
\cite{Bollobas2001}.
More specifically, the SAW path lengths between 
a given pair of nodes,
$i$ and $j$, are distributed in the range bounded from
below by the shortest path length between these nodes
\cite{Katzav2015,Nitzan2016}
and 
from above by
the longest non-overlapping path between them
\cite{Karger1997}.
From a theoretical point of view, the SAW path length corresponds
to the attrition length
\cite{Herrero2005}.
Using time units rather than length units,
we also refer to the path lengths of SAWs as
{\it last hitting times}.

In a recent paper
\cite{Tishby2016} 
we obtained analytical results for the distribution of
SAW path lengths, or
last hitting times, on 
ER
networks
\cite{Erdos1959,Erdos1960,Erdos1961}.
These 
SAW paths are often referred to as kinetic growth self-avoiding walks 
\cite{Herrero2007}, 
or true self avoiding
walks 
\cite{Slade2011}.
This is 
in contrast to the SAW paths which are uniformly sampled among
all possible self avoiding paths of a given length.
It was found that
the distribution of path lengths follows 
the Gompertz distribution
\cite{Gompertz1825,Johnson1995,Shklovskii2005,Ohishi2008}.
This means that the SAWs exhibit a termination rate per step which
increases exponentially with the number of steps already pursued.
In the limit of dilute networks it was found that the probability density
function 
of the path lengths,
$P(d=\ell)$, 
is a monotonically decreasing function and most paths
are short. 
As the connectivity of the network is increased, the paths 
become longer and the path
length distribution develops a peak.
Further increase in the connectivity shifts the peak to the right.
We derived analytical expressions for several central measures
(mean, median and mode) and for dispersion measures 
(standard deviation and the interquartile range) 
of this distribution.

Another important time scale which appears 
in random walks on networks is the
{\it first hitting time}
\cite{Debacco2015},
also referred to as the first intersection length
\cite{Herrero2003,Herrero2005b}.
This time scale emerges in a class of RW models which are not 
restricted to be self avoiding. In these models the RW keeps hopping
between adjacent nodes until it enters a node which it has already visited
before. At this point the path is terminated. 
The number of time steps up to the termination of the path, which
coincides with the path length, 
is called the first hitting time.
For a given network size, the first hitting time tends to 
increase as the network becomes
more strongly connected, because as the degree of a node 
is increased it takes longer
for the RW to visit a given fraction of its neighbors.
However, it is always much smaller than the last hitting 
time, namely the length of the corresponding
SAW path.
This is due to the fact that the RW may be terminated at any 
time step $t>1$ by randomly hopping into an
already visited node, even if the current node has one or more 
yet-unvisited neighbors, while the SAW
terminates only when the current node does not have any 
yet-unvisited neighbors.

A RW model which terminates at its first hitting time
can be cast in 
the language of foraging theory
as a model describing a wild animal, which is 
randomly foraging in a random
network environment. 
Each time the animal visits 
a node it consumes all the food
available in this node and needs to move on to 
one of the adjacent nodes.
The model describes rather harsh conditions, in 
which the regeneration of 
resources is very slow and the visited nodes do 
not replenish within the lifetime
of the forager. Moreover, the forager does not 
carry any reserves and in order
to survive it must hit a vital node each and every time. 
More realistic variants of this model have been 
studied on lattices of 
different dimensions. It was shown 
that under slow regeneration
rates, the forager is still susceptible to starvation, 
while above some threshold regeneration 
rate, the probability of starvation diminishes
\cite{Chupeau2016a}. 
The case in which the forager
carries sufficient resources that enable it to avoid starvation even when it 
visits up to $S$ non-replenished nodes in a row, was also studied 
\cite{Benichou2014,Chupeau2016}.

In this paper we present analytical results for the
distribution 
$P(d=\ell)$
of first hitting times of RWs on 
an ensemble of
ER networks.
In our analysis, we utilize the fact that up to its 
termination the RW follows an
SAW path. 
The path pursued by the RW may terminate either by 
backtracking into the previous node
or by retracing itself, namely stepping into a node which was 
already visited two or more time steps earlier.
By calculating the probabilities of these two scenarios, we
obtain analytical results for the distribution of the
first hitting times of RWs on ER networks.
The results are found to be in excellent agreement with numerical
simulations.
We obtain analytical results for the overall probabilities,
$p_b$ and $p_r$,
that a RW, starting from a random initial node,
will be terminated by backtracking or by retracing,
respectively.
It is found that in dilute networks most paths are terminated
by backtracking while in dense networks most paths are
terminated by retracing.
We also obtain expressions 
for the
conditional distributions of path lengths,
$P(d=\ell | b)$ 
and
$P(d=\ell | r)$
for the RWs which are terminated by
backtracking or by retracing, 
respectively.
These results provide useful insight into the general problem of 
survival analysis and the statistics of mortality 
or failure rates,
under conditions in which two or more 
failure mechanisms coexist
\cite{Finkelstein2008,Gavrilov2001}.

The paper is organized as follows.
In Sec. 2 we describe the random walk model 
on an ER network.
In Sec. 3 we briefly describe some properties of the ER network
which are important for the analysis presented in this paper.
In Sec. 4 we consider the temporal
evolution of two subnetworks, one consisting of the
nodes already visited by the RW and the other consisting
of the yet unvisited nodes.
In Sec. 5 we derive analytical results 
for the distribution of first hitting times
of RWs on ER networks.
In Sec. 6 we obtain analytical expressions for two central measures
(mean and median) and for a dispersion measure (the standard deviaion) 
of this distribution.
In Sec. 7 we calculate the termination probabilities
by the backtracking and by the retracing 
mechanisms.
We also calculate the conditional path length
distributions for RWs which terminate by each one of these two mechanisms.
The results are summarized and discussed in Sec. 8.

\section{The random walk model}

Consider a RW on a random network of $N$ nodes.
Each time step the walker chooses randomly one of the neighbors
of its current node, and hops to the chosen node. 
Here we study the case in which the RW path is
terminated upon the first time it steps into an already visited node.
The resulting path length, $d$,
namely the number of steps 
pursued by the RW
until its termination, 
is referred to as the first hitting time
or as the
first intersection length.
In the analysis below we do not include the termination step itself
as a part of the RW path. 
This means that the path length of a RW which pursued
$\ell$ steps and was terminated in the $\ell+1$ step, 
is $d=\ell$. 
The path includes $\ell+1$ nodes, 
since the initial node is also counted as
a part of the path.
Interestingly, the paths of the RWs up to the termination step
are, in fact, SAW paths, since each node along the path is visited
only once. The termination may take place in two possible ways.
In one scenario the path is terminated when the 
RW backtracks into the previous node,
while in the other scenario it
steps into a node which was already visited at an earlier time.

\section{The Erd\H{o}s-R\'enyi network}

In this section we briefly summarize the properties of ER 
networks which are of particular relevance to the
analysis presented below.
The ER network is the simplest model of a random network
\cite{Erdos1959,Erdos1960,Erdos1961}.
It has been studied extensively over more than five decades
and many of its properties are known exactly
\cite{Bollobas2001}.
The ER network,
denoted by
$ER(N,p)$,
consists of $N$ nodes such that each pair of nodes
is connected with probability $p$. 
The degree distribution 
of an ER network 
is a binomial distribution, $B(N,p)$.
In the limit
$N \rightarrow \infty$
and
$p \rightarrow 0$,
where the mean degree
$c=(N-1)p$
is held fixed,
it converges to a
Poisson distribution of the form

\begin{equation}
p_0(k)=\frac{{c}^{k}}{k!}e^{-c}.
\label{eq:poisson}
\end{equation}

\noindent
Clearly, there are no degree-degree correlations between adjacent nodes.
In fact, ER networks can be considered as a maximum entropy 
ensemble under the constraint that the mean degree is fixed.
In the asymptotic limit 
($N \rightarrow \infty$),
the ER
network exhibits a phase transition at 
$c=1$ (a percolation transition), such that for
$c<1$
the network consists only of small clusters 
and isolated nodes, while for 
$c>1$
there is a giant cluster which includes 
a macroscopic fraction of the network, in addition
to the small clusters and isolated nodes. 
At a higher value of the connectivity, namely at 
$c = \ln N$, 
there is a second transition, above which 
the entire network is included in
the giant cluster and there are 
no isolated components. 
For intermediate values of $c$, namely for
$1 < c < \ln N$, the fraction, $g=g(c)$, of nodes which belong 
to the giant cluster
is given by the implicit equation
$g = 1 - \exp(-cg)$.
Solving for $g$, one obtains
$g(c)= 1 + {W(-c e^{-c})}/{c}$,
where $W(x)$ is the Lambert $W$ function.
Thus, the fraction of nodes which belong to network components
apart from the giant cluster
is given by $1-g$. 
This includes nodes which reside on
small clusters as well as isolated nodes.
The fraction, 
$i=i(c)$, 
of isolated nodes among all nodes 
in the network is given by
$i(c) = \exp(-c)$.
Thus, the fraction, 
$h=h(c)$, 
of nodes which reside on isolated clusters of size
$s>1$ 
is given by
$h(c) = 1 - g(c) - i(c)$,
or more explicitly by
$h(c) = - {W(-c e^{-c})}/{c} - e^{-c}$.
Here we focus on the regime
above the percolation transition, namely 
$c>1$. 
In order to avoid the trivial case of a RW starting
on an isolated node, we performed the analysis presented
below for the case in which the initial node is non-isolated.
However, isolated clusters which consist of two
or more nodes are not excluded.
Thus, the results presented below correspond to the entire 
network rather than to the giant component alone.

\section{Evolution of the subnetworks of visited and the yet-unvisited nodes}

Consider an $ER(N,p)$ network.
The degree $k_i$ of node $i=1,\dots,N$ 
is the number of links connected to this node.
The RW divides the network into two sub-networks:
one consists of the already visited nodes and the
other consists of the yet unvisited nodes. 
After $t$ time steps the size of the subnetwork of 
visited nodes is $t+1$ 
(including the initial node),
while the size of the network of
yet unvisited nodes is
$N(t)=N-t-1$. 
The degree distributions of both sub-networks evolve in time.
We denote the degree distribution of the sub-network of the
yet unvisited nodes at time $t$ by
$p_t(k)$, $k=0,\dots,N(t)-1$,
where
$p_0(k)$,
is the original degree distribution
given by Eq.
(\ref{eq:poisson}).
The mean degree,
$c(t) = \langle k \rangle_t$,
of this sub-network, which is given by

\begin{equation}
c(t) = \sum_{k=0}^{N(t)-1} k p_t(k),
\label{eq:<k>}
\end{equation}

\noindent
evolves accordingly,
where $c(0)=c$.

For random walks on random networks,
there is a higher probability for the walker to
visit nodes with high degrees. 
More precisely, 
the probability that in the next time step the RW will visit a node 
of degree $k$
is 
$k p_0(k)/c$.
Conditioned on stepping into one of the yet 
unvisited
nodes adjacent to the current node,
the probability of stepping into a node of degree $k$ 
is 
$k p_t(k)/c(t)$.
In Ref.
\cite{Tishby2016}
it was shown that for an SAW on an ER network
the degree distribution of the
subnetwork of the yet unvisited nodes,
at time $t$, is

\begin{equation}
p_t(k) = \frac{c(t)^k}{k!}e^{-c(t)},
\label{eq:p_t(k)}
\end{equation}

\noindent
where 

\begin{equation}
c(t) = c - pt
\label{eq:coft}
\end{equation}

\noindent
is the mean degree of this subnetwork.
These exact results imply
that the subnetwork of the yet unvisited nodes
remains an ER network, while its mean degree decreases linearly
in time.
A special property of the 
Poisson distribution is that
$k p_t(k) / c(t) = p_t(k-1)$.
This means that,
the probability that the node
visited at time $t+1$ will be of  
degree $k$ 
is given by 
$p_t(k-1)$. 
Since the RW follows an SAW path until it terminates,
this result applies also the RW model studied here.

\section{The distribution of first hitting times}

Consider a RW on an ER network, 
which starts from a random node with degree
$k \ge 1$ (non-isolated node).
The RW hops randomly between nearest neighbor nodes.
It continues to hop as long as all the nodes it steps 
into have not been visited before.
Once the RW steps into a node which has already been visited,
the path is terminated.
We distinguish between two termination scenarios.
In the first scenario, the RW hops back into the previous
node (backtracking step).
In the second scenario 
the RW hops into a node which was already visited at an earlier time
(retracing step).

In case that the RW has pursued $t$ steps, without visiting any 
node more than once, the path length is guaranteed to be $d > t-1$.
At this point, the probability that the path will not
be terminated in the $t+1$ step is denoted by the conditional probability
$P(d>t|d>t-1)$. This conditional probability can be expressed as a product
of the form

\begin{equation} 
P(d > t|d > t-1) =  P_b(d > t|d > t-1) P_r(d > t|d > t-1).  
\label{eq:cond3}
\end{equation}

\noindent
The conditional probability
$P_b(d > t|d > t-1)$ 
is the probability that the RW will 
not backtrack its path at the $t+1$ time step and will thus avoid the
first termination scenario.
Given that the RW has not backtracked its path, 
the conditional probability
$P_r(d > t|d > t-1)$ 
is the probability that it will 
not step into a node already visited at an earlier time, thus
avoiding the 
second termination scenario.

An RW which at time $t$ resides in a node of degree $k$ may
hop in the next time step to each one of its $k$ neighbors.
One of these neighbors is the previous node, visited by the RW at time $t-1$.
Thus, the probability of a backtracking step into the previous node is
$1/k$.
The degree distribution of nodes visited by an RW is given by
$k p_0(k)/c$. 
Thus the probability of backtracking is

\begin{equation}
\left\langle  \frac{1}{k}  \right\rangle =
\sum_{k=1}^{N-1} \frac{1}{k} \frac{k p_0(k)}{c}.
\end{equation}

\noindent
Evaluation of the right hand side yields

\begin{equation}
\left\langle  \frac{1}{k}  \right\rangle =
\frac{1-e^{-c}}{c}.
\end{equation}

\noindent
Thus, the probability that the RW will not backtrack its path at time
$t+1$ is given by

\begin{equation}
P_b(d>t|d>t-1) = 1 - \left( \frac{1-e^{-c}}{c} \right).
\end{equation}

\noindent
Note that this probability does not depend on $t$.

\begin{figure}
\centerline{
\includegraphics[width=7cm]{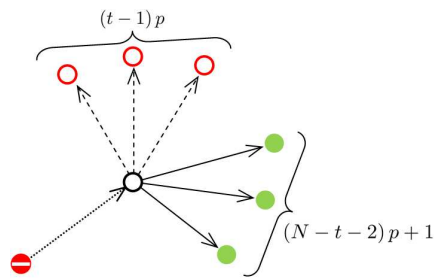}
}
\caption{
The evaluation of the probability, 
$P_r(d>t | d>t-1)$,
that an RW will not terminate by retracing at the $t+1$ time step
is illustrated. 
The total number of neighbors of the current node, 
apart from the previous node, is
$(N-3)p + 1$, of which $(N-t-2)p + 1$ 
have not been visited before.
} 
\label{fig:1}
\end{figure}

Provided that the RW was not terminated by backtracking at the
$t+1$ step, we will now evaluate the probability
$P_r(d > t|d > t-1)$ 
that it will also not be terminated by retracing
its path in that step.
Apart from the current node and the previous node, there are $N-2$
possible nodes which may be connected to the current node, 
with probability $p$
(Fig. \ref{fig:1}).
The fact that the possibility of backtracking 
was already eliminated for the $t+1$ step,
guarantees that at least one of these $N-2$ 
nodes is connected to the current node
with probability $1$
(otherwise, the only possible move would have been to hop back to the previous node).
This leaves $N-3$ nodes which are connected to the
current node with probability $p$.
Thus, the expectation value of the number of neighbors of the current
node, to which the RW may hop in the $t+1$ step, is $(N-3)p+1$.
Due to the local tree-like structure of ER networks, 
it is extremely unlikely that
the one node which is guaranteed to be connected to the current node has
already been visited. 
This is due to the fact that the path from such earlier
visit all the way to the current node is essentially a loop. 
Therefore, we
conclude that this adjacent node has not yet been visited.
Since the number of yet unvisited nodes is $N-t-1$ we conclude that
the current node is expected to have
$(N-t-2)p+1$ neighbors which have not yet been visited.
As a result, the probability that the RW will hop into one of the
yet-unvisited nodes is given by

\begin{equation}
P_r(d>t | d> t-1) = \frac{(N-t-2)p+1}{(N-3)p+1}.
\label{eq:P_r0}
\end{equation}

\noindent
Inserting 
$c=(N-1)p$
and 
$c(t)=(N-t-1)p$
we obtain

\begin{equation} 
P_r(d>t|d>t-1) = \frac{c(t)-p+1}{c-2p+1}. 
\label{eq:P_r2}
\end{equation}

\noindent
In the asymptotic limit this expression can be
approximated by

\begin{equation} 
P_r(d>t|d>t-1) = \frac{c(t)+1}{c+1}. 
\label{eq:P_r}
\end{equation}

\noindent
Combining the results presented above, it is found that
the probability that the RW will proceed from time
$t$ to time $t+1$ 
is given by the conditional probability

\begin{equation} 
P(d > t|d > t-1) =  
\left(  \frac{c - 1 + e^{-c} }{c}  \right)
\left[ \frac{c(t)-p+1}{c-2p+1} \right].
\label{eq:cond3p}
\end{equation}

\begin{figure}
\centerline{
\includegraphics[width=7cm]{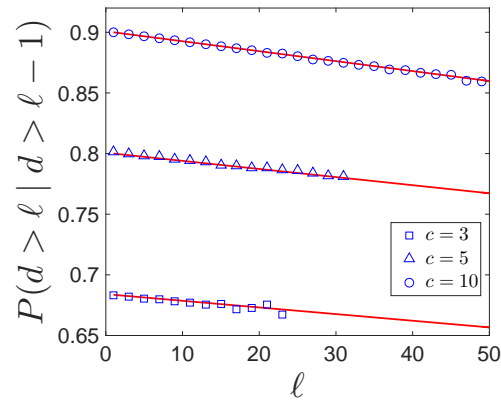}
}
\caption{
The conditional probability 
$P(d > \ell|d > \ell-1)$ 
vs. $\ell$,
obtained from Eq.
(\ref{eq:cond3p})
(solid lines)
and 
from numerical simulations of RWs 
(symbols)
on ER networks of
size $N=1000$ and mean degrees $c=3$, $5$ and $10$ 
(squares, triangles and circles, respectively).
The analytical and numerical results are found to
be in good agreement.
} 
\label{fig:2}
\end{figure}

\noindent
In Fig. 
\ref{fig:2} 
we present
the conditional probability 
$P(d > t|d > t-1)$ 
vs. $t$ for a network 
of size $N=1000$ and for three values of $c$. 
The analytical results (solid lines) 
obtained from Eq.
(\ref{eq:cond3p})
are found to be in good agreement with numerical simulations
(symbols),
confirming the validity of this equation.
Note that the numerical results become more noisy as $t$
increases, due to diminishing statistics, 
and eventually terminate.
This is particularly apparent for the smaller values of $c$.

The probability that the path length of the RW will
be longer than $\ell$ is given by

\begin{equation} 
P(d>\ell) = P(d>0) \prod_{t=1}^{\ell} P(d > t|d > t-1),
\label{eq:cond}
\end{equation}

\noindent
where $P(d>0)=1$,
since the initial node is not isolated.
Using Eq.
(\ref{eq:cond3})
the probability $P(d>\ell)$ can be written as a product of the form

\begin{equation} 
P(d>\ell) =  P_b(d>\ell) P_r(d>\ell),
\label{eq:cond2}
\end{equation}

\noindent
where

\begin{equation} 
P_b(d>\ell) =  \prod_{t=1}^{\ell} 
\left( \frac{c - 1 + e^{-c} }{c} \right),
\label{eq:cond4}
\end{equation}

\noindent
and

\begin{equation} 
P_r(d>\ell) =  \prod_{t=2}^{\ell} 
\left[ \frac{c(t)-p+1}{c-2p+1} \right].
\label{eq:cond6}
\end{equation}

%

\noindent
The calculation of the tail distribution
$P_b(d>\ell)$ 
is simplified by the fact that 
$P_b(d>t | d>t-1)$
does not depend on $t$.
As a result, Eq.
(\ref{eq:cond4})
can be written in the form

\begin{equation}
P_b(d>\ell) = e^{- \beta \ell},
\label{eq:P_b1}
\end{equation}

\noindent
where

\begin{equation}
\beta = \ln \left(\frac{c}{c-1+e^{-c}}\right).
\label{eq:beta}
\end{equation}

%

\noindent
Taking the logarithm of $P_r(d>\ell)$,
as expressed in Eq.
(\ref{eq:cond6}),
we obtain

\begin{equation}
\ln \left[P_r(d>\ell)\right] = 
\sum_{t=2}^{\ell} \ln \left[ \frac{c(t)-p+1}{c-2p+1} \right].
\label{eq:P_rsum}
\end{equation}

\noindent
Replacing the sum by an integral 
and plugging in the expression for $c(t)$ 
from Eq.
(\ref{eq:coft}),
we obtain

\begin{equation}
\ln [P_r(d>\ell)]
\simeq
\int_{3/2}^{\ell+1/2}
\ln \left[1-\frac{c\cdot\left(t-1\right)}{\left(N-1\right)
\left(c-2p+1\right)}\right]dt.
\label{eq:P_rint}
\end{equation}

\noindent
The integrand can be simplified by replacing
$N-1$ by $N$ and $c-2p+1$ by $c+1$,
which is accurate when $p \ll 1$.
Using the notation

\begin{equation}
\alpha=\sqrt{ \frac{N(c+1)}{c} },
\end{equation}

\noindent
and solving the integral, we obtain

\begin{eqnarray}
\ln[P_r(d>\ell)] &\simeq&
\left( \ell - \frac{1}{2} - \alpha^2 \right)
\ln\left[ 1 - \frac{ \left( \ell - \frac{1}{2} \right)}{\alpha^2} \right]-\ell 
\\ \nonumber
&+&
\left(\alpha^2 - \frac{1}{2} \right) \ln \left( 1 - \frac{1}{2\alpha^2} \right) + 1.
\label{eq:P_r4}
\end{eqnarray}

\noindent
In the approximation of the sum of 
Eq. (\ref{eq:P_rsum}) by the integral of Eq. (\ref{eq:P_rint})
we have used the
formulation of the middle Riemann sum. Since the function
$\ln[P_r(d>\ell)]$ is a monotonically decreasing function, the value of
the integral is over-estimated by the left Riemann sum, $L_{\alpha}(\ell)$, and under-estimated
by the right Riemann sum, $R_{\alpha}(\ell)$. The error involved in this approximation is thus 
bounded by the difference $\Delta_{\alpha}(\ell) = L_{\alpha}(\ell) - R_{\alpha}(\ell)$,
which satisfies
$\Delta_{\alpha}(\ell) < \ln(1 - \ell/\alpha^2)$.
Thus, the relative error in $P(d>\ell)$ due to the approximation of the sum
by an integral is bounded by
$\eta_{\rm SI} = \ell/\alpha^2$,
which scales like $\ell/N$. 
Comparing the values obtained from the sum and the integral
over a broad range of parameters, we find that the pre-factor of
the error is very small, so in practice the error introduced by approximation of the
sum by an integral is negligible.

Under the assumption that the RW paths are short 
compared to the network size, namely
$\ell \ll N$, 
one can use the approximation

\begin{equation}
\ln\left(1-\frac{\ell}{\alpha^2}\right)
\simeq
-\frac{\ell}{\alpha^2} 
- \frac{\ell^2}{2 \alpha^4}
+ O\left( \frac{\ell^3}{\alpha^6} \right).
\label{eq:lnapprox}
\end{equation}

\noindent
Plugging this approximation into Eq.
(\ref{eq:P_r4})
yields

\begin{equation}
P_r\left(d>\ell\right)
\simeq
\exp \left[-\frac{\ell (\ell - 1)}{2 \alpha^2}\right].
\label{eq:tail_simp2}
\end{equation}

\noindent
Combining the results obtained above for
$P_b(d>\ell)$
[Eq. (\ref{eq:P_b1})]
and for
$P_r(d>\ell)$
[Eq. (\ref{eq:tail_simp2})]
we obtain
%
%

\begin{equation}
P\left(d>\ell\right)
\simeq
\exp \left[ - \frac{\ell(\ell-1)}{2\alpha^2}
- \beta \ell \right].
\label{eq:tail_simp3}
\end{equation}

\noindent
Thus, the distribution of path lengths is a product of 
an exponential distribution and a
Rayleigh distribution,
which is a special case of the Weibull distribution
\cite{Papoulis2002}.
Considering the next order in the series expansion of Eq.
(\ref{eq:lnapprox})
we find that the relative error in 
Eq. (\ref{eq:tail_simp3})
for $P(d>\ell)$
due to the truncation of the Taylor expansion after the second order
is $\eta_{\rm TE} =  \ell^3/(6 \alpha^4)$,
which scales like $\ell^3/N^2$.
This error is very small as long as $\ell \ll N^{1/2}$.
Note that paths of length $\ell \simeq N^{1/2}$,
for which the error 
in $P(d>\ell)$
is noticeable, become prevalent only
in the limit of dense networks, where $c > N^{1/2}$.

\begin{figure}
\centerline{
\includegraphics[width=12cm]{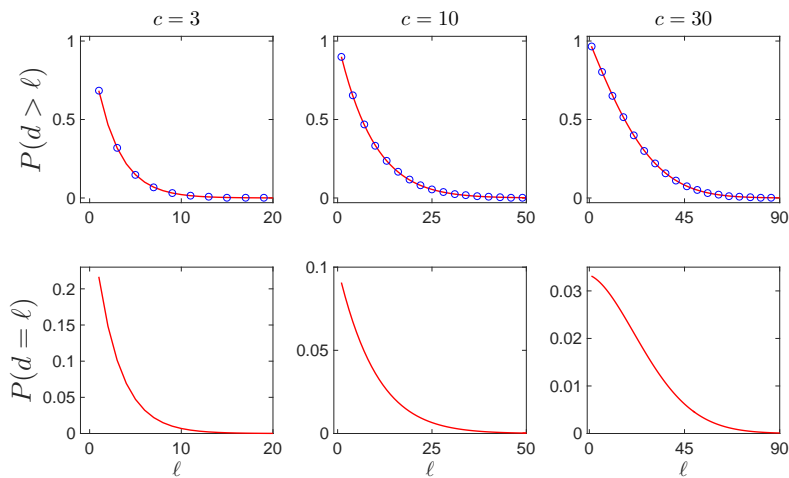}
}
\caption{
The tail distributions 
$P(d > \ell)$ vs. $\ell$
of the first hitting times
of RWs for ER networks
of size $N=1000$ 
and 
$c=3$, $10$ and $30$.
The theoretical results, obtained from Eq.
(\ref{eq:tail_simp3}) 
(solid lines) and the results obtained
from numerical simulations (circles)
are shown in the top row, and
are found to be in excellent agreement with each other.
The corresponding probability density functions,
$P(d=\ell)$,
obtained from Eq.  
(\ref{eq:pdf}),
are shown in the bottom row. The agreement with the
numerical results is already established in the top row and therefore the numerical
data is not shown in the bottom row.
}
\label{fig:3}
\end{figure}

In Fig.
\ref{fig:3}
we present the tail distributions
$P(d>\ell)$
vs.
$\ell$
of the first hitting times of RWs on ER networks of size
$N=1000$ and 
mean degrees 
$c=3$, $10$ and $30$.
The theoretical results (solid lines),
obtained from
Eq.
(\ref{eq:tail_simp3}),
are found to be in excellent agreement 
with the numerical simulations
(symbols).
The probability density function 
$P(d=\ell)$
is given by 

\begin{equation}
P\left(d=\ell\right) = P\left(d>\ell-1\right)-P\left(d>\ell\right).
\label{eq:pdf}
\end{equation}

\section{Central and dispersion measures of the path length distribution}

In order to characterize the distribution of 
first hitting times of RWs on ER networks we
derive expressions for the mean and median of 
this distribution.
The mean of the distribution can be obtained
from the tail-sum formula



\begin{equation}
\ell_{\rm mean}(N,c) =
1 + \sum_{\ell=1}^{N-2} P(d>\ell), 
\label{eq:tailsum2}
\end{equation}

\noindent
under the assumption that the initial node is a non-isolated node.
Expressing the sum 
as an integral we obtain

\begin{equation}
\ell_{\rm mean}(N,c) =
1 + \int_{1/2}^{N-3/2} P(d>\ell)d\ell,
\label{eq:ell_mean1}
\end{equation}

\noindent
where the range of integration is shifted downwards by $1/2$,
such that the summation over each integer, $i$, is replaced by an
integral over the range $(i-1/2,i+1/2)$.
Inserting 
$P(d>\ell)$
from Eq.
(\ref{eq:tail_simp3})
and solving the integral, we obtain

\begin{equation}
\ell_{\rm mean}
=1+\sqrt{ \frac{\pi}{2}}\alpha e^{\frac{ ( \alpha^{2} \beta - 1 ) \beta }{2}}
\left[{\rm erf}
\left(
\frac{ \alpha^2 \beta + N - 2 }{\sqrt{2}\alpha}
\right)
-{\rm erf}\left( \frac{\alpha \beta}{ \sqrt{2} } \right)
\right].
\label{eq:ell_mean4}
\end{equation}

\noindent
Using the relative error of
Eq. (\ref{eq:tail_simp3})
for $P(d>\ell)$,
we estimate the relative error of $\ell_{mean}$
by 
$\eta =  \ell_{\rm mean}^3/(6 \alpha^4)$,
which scales like 
$\ell_{\rm mean}^3/N^2$.
We can safely approximate the first 
${\rm erf}$ 
on the right hand side of Eq.
(\ref{eq:ell_mean4})
to be equal to $1$, 
and obtain

\begin{equation}
\ell_{\rm mean}
\simeq
1+\sqrt{\frac{\pi}{2}}\alpha e^{ \frac{ ( \alpha^{2} \beta - 1 ) \beta }{2}}
\left[1-{\rm erf} \left(\frac{\alpha \beta}{\sqrt{2}}\right)\right].
\label{eq:ell_mean5}
\end{equation}

\noindent
In the limit of dense networks, where

\begin{equation}
c > c^{\ast} = \sqrt{ \frac{2}{\pi} } \sqrt{N},
\end{equation}

\noindent
$\ell_{\rm mean}$ can be expressed in the asymptotic form

\begin{equation}
\ell_{\rm mean} \simeq 1 + \sqrt{ \frac{\pi}{2} } \sqrt{N} 
\left[ 1 - \sqrt{ \frac{2}{\pi} } \frac{ \sqrt{N} }{c} 
+ \frac{1}{2} \frac{N}{c^2} 
+ O \left( \frac{N^{\frac{3}{2}}}{c^3} \right)
\right].
\end{equation}

\begin{figure}
\centerline{
\includegraphics[width=7cm]{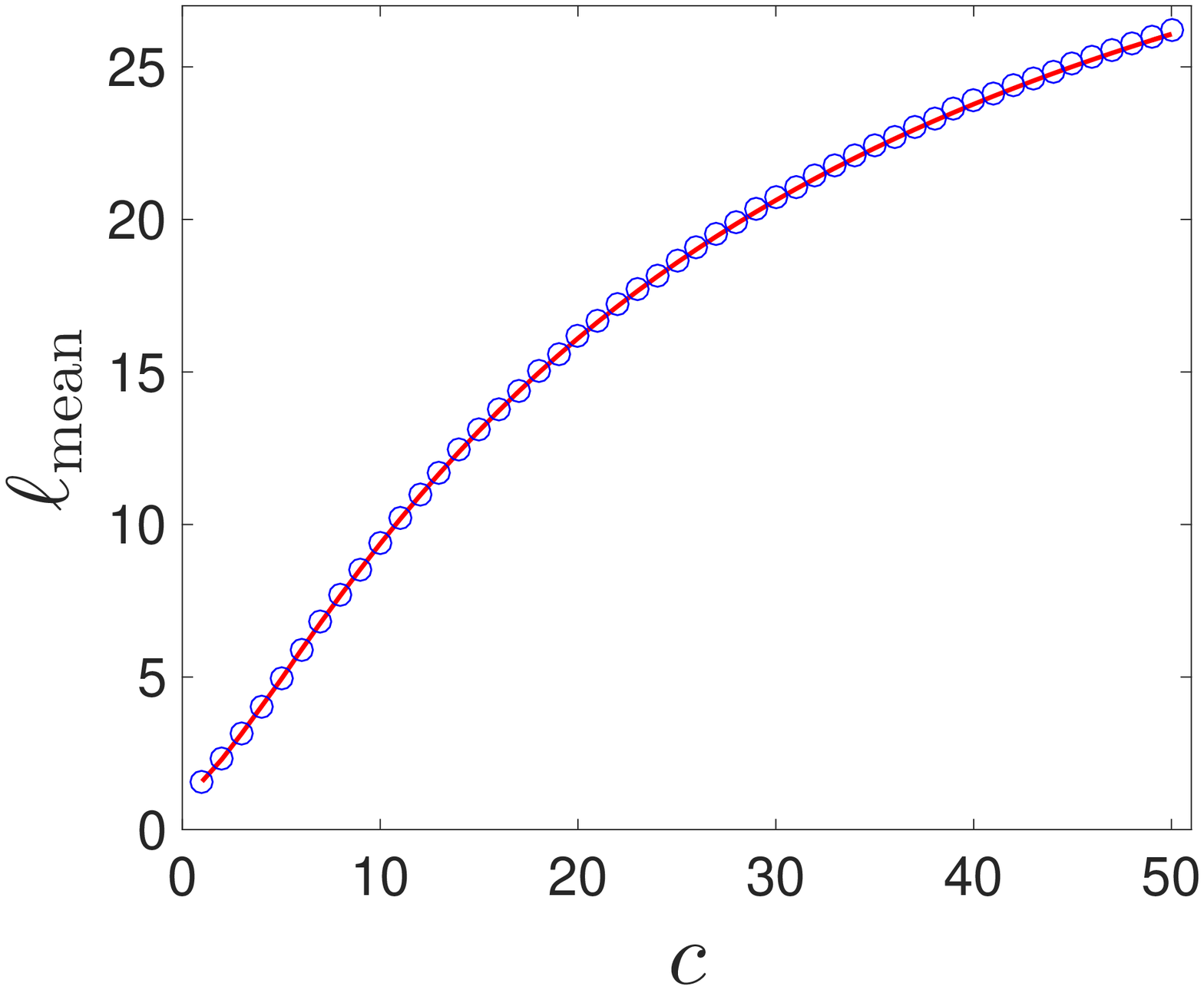}
}
\caption{
The mean of the distribution of first hitting times,
$\ell_{\rm mean}$, as a function of the mean degree,
$c$, for ER networks of size
$N=1000$.
The analytical results (solid line), 
obtained from Eq.
(\ref{eq:ell_mean5})
are in excellent agreement with numerical 
simulations 
(circles). 
}
\label{fig:4}
\end{figure}

\noindent
In Fig.
\ref{fig:4}
we present the mean value, $\ell_{\rm mean}$
of the distribution of first hitting times
as a function of the mean degree $c$,
for ER networks of size $N=1000$.
The agreement between the theoretical results,
obtained from Eq.
(\ref{eq:ell_mean5})
and the numerical simulations is very good for all values of
$c$.

\begin{figure}
\centerline{
\includegraphics[width=7cm]{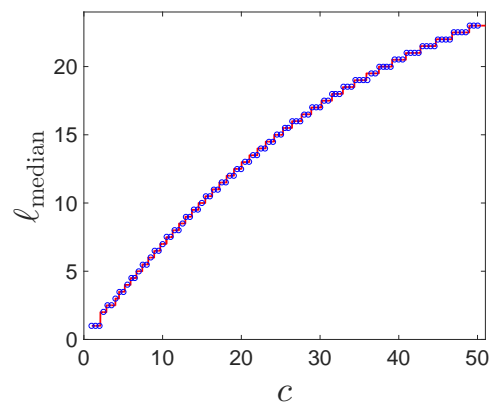}
}
\caption{
The median of the distribution of first hitting times,
$\ell_{\rm median}$, as a function of the mean degree,
$c$, for ER networks of size
$N=1000$.
The analytical results (solid line) 
are in excellent agreement with numerical 
simulations 
(circles). 
}
\label{fig:5}
\end{figure}

To obtain a more complete characterization of the distribution of first
hitting times, it is also useful to evaluate its median, 
$\ell_{\rm median}$.
Here the
median is defined as the value of $\ell$ for which
$| P(d>\ell) - P(d<\ell) | \rightarrow {\rm min}$,
where $\ell$ may take either an integer or a half-integer value.
In Fig.
\ref{fig:5}
we present the median, 
$\ell_{\rm median}$,
of the distribution of first hitting times
as a function of the mean degree, $c$,
for ER networks of size $N=1000$.
The agreement between the theoretical results
and the numerical simulations is 
very good for all values of $c$.
Note that in the evaluation of $\ell_{\rm median}$ we use the
accurate expression of $P_r(d>\ell)$, given by Eq.
(\ref{eq:P_r4}) rather than the approximate expression of 
Eq. (\ref{eq:tail_simp2}).
Using the approximate expression gives rise to small
discrepancies in the locations of the edges of the steps
for large values of $c$.

In the limit of very high connectivity,
Eq. 
(\ref{eq:ell_mean5})
can be approximated by
$\ell_{\rm mean} \simeq 1 + ( \pi  N/2 )^{1/2}$.
Thus, in such dense networks, the mean path
length becomes independent of the mean degree $c$,
and scales according to 
$\ell_{\rm mean} \sim \sqrt{N}$.
This can be understood as follows.
In this limit, the backtracking probability 
is very low and thus the backtracking-induced
termination of paths is no longer of much significance.
Instead, retracing becomes the main reason for termination of paths. 
Due to the very high connectivity, the hopping between adjacent nodes
can be considered as the simple combinatorial problem of randomly choosing 
one node at a time from a set of $N$ nodes, 
allowing each node to be chosen more than once.
The probability that such process will yield $\ell$ distinct nodes is
given by
$P(d>\ell-2) \simeq \exp( - \ell^2/N )$.
Thus, in this limit the median is given by
$\ell_{\rm median} \simeq \sqrt{N \ln 2}$.
This result is analogous to the birthday problem, where
$N=365$ and 
$\ell_{\rm median}$
is the smallest number of participants in a party such that with probability 
of at least $1/2$ there is at least one pair that has the same birthday
\cite{Bloom1973}.

The moments of the distribution of RW path lengths,
$\langle \ell^n \rangle$,
are given by
the tail-sum formula
\cite{Pitman1993}

\begin{equation}
\langle \ell^n \rangle = 
\sum_{\ell=0}^{N-1} [(\ell+1)^n - \ell^n] P(d>\ell).
\label{eq:tail_sumn}
\end{equation}

\noindent
Using this formula to evaluate the second moment 
and replacing the sum by an integral we obtain

\begin{equation}
\langle \ell^{2} \rangle = 1 +
\int_{\frac{1}{2}}^{N-\frac{1}{2}}\left(2\ell+1\right)
\exp\left[-\frac{\ell(\ell-1)}{2 \alpha^{2} } - \beta \ell \right] d\ell.
\label{eq:ell2}
\end{equation}

\noindent
Solving the integral and taking the large network limit,
we obtain

\begin{equation}
\langle \ell^2 \rangle
\simeq
1 + 
2 \alpha^2 e^{ - \frac{\beta}{2} }
+
\sqrt{2 \pi} {\alpha}
\left(1 - \alpha^{2}\beta \right)
e^{\frac{ ( \alpha^2 \beta - 1 ) \beta}{2}}
\left[1 - {\rm erf} \left(\frac{\alpha \beta}{\sqrt{2} }\right)\right].
\label{eq:ell_sqr}
\end{equation}

\noindent
The standard deviation of the distribution of path lengths is
thus given by

\begin{equation}
\sigma_{\ell}^2(c) = \langle \ell^2 \rangle - \ell_{mean}^{2},
\label{eq:sigmaell}
\end{equation}

\noindent
where $\langle \ell^2 \rangle$
is given by Eq.
(\ref{eq:ell_sqr})
and
$\ell_{mean}$
is given by Eq.
(\ref{eq:ell_mean5}).
In the dense network limit,
where $c > c^{\ast}$,
the standard deviation 
can be approximated by

\begin{equation}
\sigma_{\ell}(c) 
\simeq 
\sqrt{\frac{4-\pi}{2}}\sqrt{N} 
\left[ 
1+ \frac{\pi - 2}{2 c(4-\pi)} 
-   
\left( \frac{\pi-2}{4-\pi} \right)
\left( \frac{N}{2c^2} \right)
+ O\left(\frac{N^2}{c^3}\right) 
\right].
\end{equation}

\begin{figure}
\centerline{
\includegraphics[width=7cm]{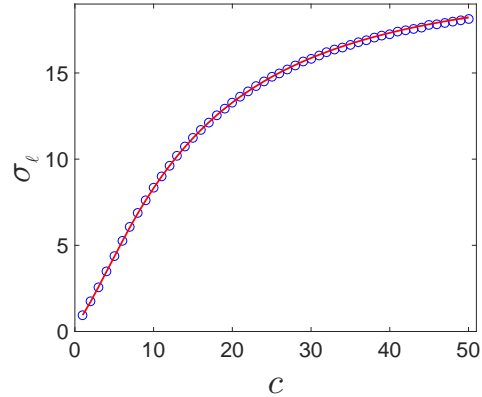}
}
\caption{
The standard deviation of the distribution of first hitting times,
$\sigma_{\ell}$, as a function of the mean degree,
$c$, for ER networks of size
$N=1000$.
The analytical results (solid line), 
obtained from Eq.
(\ref{eq:sigmaell})
are in excellent agreement with numerical 
simulations 
(circles). 
}
\label{fig:6}
\end{figure}

\noindent
In Fig.
\ref{fig:6}
we present the standard deviation, 
$\sigma_{\ell}(c)$
of the distribution of first hitting times
as a function of the mean degree, $c$,
for ER networks of size $N=1000$.
The agreement between the theoretical results,
obtained from Eq. 
(\ref{eq:sigmaell}),
and the numerical simulations is very good for all values of
$c$.

\section{Analysis of the two termination mechanisms}

The RW model studied here may terminate either due
to backtracking or due to retracing its path.
The backtracking mechanism may occur starting from the
second step of the RW. 
The expected probability of backtracking
is $[1-\exp(-c)]/c$ 
at any step afterwards, regardless of the number
of steps already pursued.
In case of termination by retracing 
the RW path forms a loop
which starts at the first visit to the termination node and ends in
the second visit.
Termination due to retracing may play a role starting from
the third step of the RW. 
The probability that the RW will terminate
due to retracing increases in time.
This is due to the fact that each visited node
becomes a potential trap.
It is thus expected that paths that terminate after a small number of
steps are most likely to be terminated by backtracking, while paths
which survive for a long time are more likely to be terminated by
retracing. Below we present a detailed analysis of the probabilities
of a RW to terminate by backtracking or by retracing. 
We also present the  
dependence of these probabilities 
on the number of steps already
pursued.

Consider a RW on an ER network, starting from a random, non-isolated node.
The RW will follow a path visiting a new node at each
one of the first $\ell$ steps.
It will terminate 
at the $\ell+1$ step, by entering an 
already visited node.
Since the failed termination step is not counted as a part of the
path, the path length in this case will be $d=\ell$.
The probability distribution function of the RW path lengths,
$P(d=\ell)$, 
is given by Eq.
(\ref{eq:pdf}).
We denote by $p_b$
the probability that a RW 
starting from a random initial node
will terminate
by backtracking 
and by $p_r$ the probability that it will terminate
by retracing.
Since these are the only two termination mechanisms in the model,
the two probabilities must satisfy
$p_b + p_r = 1$.

While the overall distributions of path lengths is given by
$P(d=\ell)$, one expects the distribution $P(d=\ell | b)$ of 
paths terminated by backtracking to differ from the distribution
$P(d=\ell | r)$ of paths terminated by retracing.
These conditional probability distributions 
are normalized,
namely they satisfy
$\sum_{t=2}^{N-1} P(d= t | b) =1$
and
$\sum_{t=3}^{N-1} P(d= t | r) =1$.
\label{eq:normr}
The overall distribution of path lengths can be expressed in terms of 
the conditional distributions according to

\begin{equation}  
P(d=\ell) = p_b P(d=\ell | b) + p_r P(d=\ell | r).
\label{eq:br}
\end{equation}

\noindent
The first term on the right hand side of Eq.
(\ref{eq:br}) 
can be written as

\begin{equation}
p_b P(d=\ell | b) =
P(d>\ell-1) \left[1-P_b(d>\ell|d>\ell-1) \right],
\label{eq:p_b}
\end{equation}

\noindent
namely as the probability that the RW will pursue $\ell$ steps
and will terminate in the $\ell+1$ step due to backtracking.
The second term on the right hand side of Eq.
(\ref{eq:br}) 
can be written as

\begin{equation}
p_r P(d=\ell | r) =
P(d>\ell-1) P_b(d>\ell|d>\ell-1)
\left[1-P_r(d>\ell|d>\ell-1)\right],
\label{eq:p_r}
\end{equation}

\noindent
namely as the probability that the RW will pursue $\ell$ steps,
then in the $\ell+1$ step it will not backtrack its path but will
retrace it by stepping into a node visited at least two steps earlier.

Summing up both sides of Eq.
(\ref{eq:p_b})
over all integer values of $\ell$ we obtain

\begin{equation}
p_b =  \left( \frac{ 1-e^{-c} }{c} \right) \sum_{\ell=1}^{N-1} P(d>\ell-1).
\end{equation}

\noindent
Using the tail-sum formula
we find that the
probability that the RW will terminate 
by backtracking is

\begin{equation}
p_b =  \left( \frac{1 - e^{-c}}{c} \right) \ell_{\rm mean}.
\label{eq:p_b2}
\end{equation}

\noindent
As a result, the probability of the RW to be terminated by
retracing its path is

\begin{equation}
p_r = 1 -  \left( \frac{1 - e^{-c}}{c} \right) \ell_{\rm mean}.
\label{eq:p_r2}
\end{equation}

\noindent
Using Eq. 
(\ref{eq:p_b})
the conditional probability
$P(d=\ell | b)$ 
can be written in the form

\begin{equation}
P(d=\ell | b) = \frac{P(d>\ell-1)}{\ell_{\rm mean}}.
\label{eq:p_lbeq}
\end{equation}

\noindent
Similarly, the conditional probability
$P(d=\ell | r)$
takes the form

\begin{equation}
P(d=\ell | r) =  \left( \frac{c-1+e^{-c}}{c+1} \right) 
\left[ \frac{c - c(\ell)}{c - (1-e^{-c}) \ell_{\rm mean}} \right] P(d>\ell-1), 
\label{eq:p_lreq}
\end{equation}

\noindent
where $c(\ell)$ is given by Eq.
(\ref{eq:coft}).
The corresponding tail distributions
take the form

\begin{equation}
P(d > \ell | b) = 
\frac{\sum_{t=\ell+1}^{N-1} P(d>t-1)}{\ell_{\rm mean}},
\label{eq:p_lbgt}
\end{equation}

\noindent
and

\begin{equation}
P(d > \ell | r) =  \left( \frac{c-1+e^{-c}}{c+1} \right) 
\sum_{t=\ell+1}^{N-1} 
\left[ \frac{c - c(t)}{c-(1-e^{-c})\ell_{\rm mean}} \right] P(d>t-1).
\label{eq:p_lrgt}
\end{equation}

\noindent
Given that a RW path was terminated after $\ell$ steps, it is of great interest
to evaluate the conditional probabilities 
$P(b | d=\ell)$ 
and
$P(r | d=\ell)$,
that the termination was caused by backtracking or by
retracing, respectively.
Using Bayes' theorem,
these probabilities can be expressed by

\begin{equation}
P(b | d=\ell) = \frac{p_b P(d=\ell | b)}{P(d=\ell)}
\end{equation}

\noindent
and

\begin{equation}
P(r | d=\ell) = \frac{p_r P(d=\ell | r)}{P(d=\ell)}.
\end{equation}

\noindent
Clearly, these distributions satisfy
$P(b | d=\ell) + P(r | d=\ell) =1$.
Inserting the conditional probabilities
$P(d=\ell | b)$
and
$P(d=\ell | r)$
from Eqs.
(\ref{eq:p_lbeq})
and
(\ref{eq:p_lreq}),
respectively, we find that

\begin{equation}
P(b | d=\ell) = 
\left( \frac{1-e^{-c}}{c} \right) 
\frac{P(d>\ell-1)}{P(d=\ell)}
\label{eq:b_ell}
\end{equation}

\noindent
and

\begin{equation}
P(r | d=\ell) = \left( \frac{c-1+e^{-c}}{c+1} \right)
\left[ 1 - \frac{c(\ell)}{c} \right] \frac{P(d>\ell-1)}{P(d=\ell)}.
\label{eq:b_ell2}
\end{equation}

\noindent
The corresponding tail distributions can be expressed in the form

\begin{equation}
P(b | d>\ell) = 
\left( \frac{1-e^{-c}}{c} \right) 
\frac{\sum_{t=\ell+1}^{N-1} P(d>t-1)}{ P(d>\ell)}
\label{eq:b_ell3}
\end{equation}

\noindent
and

\begin{equation}
P(r | d>\ell) = 
\left( \frac{c-1+e^{-c}}{c+1} \right) 
\sum_{t=\ell+1}^{N-1} 
\left[ 1 - \frac{c(t)}{c} \right] \frac{P(d>t-1)}{P(d>\ell)}.
\label{eq:r_ell}
\end{equation}

\noindent
These distributions also satisfy
$P(b | d>\ell) + P(r | d>\ell) =1$.

\begin{figure}
\centerline{
\includegraphics[width=7cm]{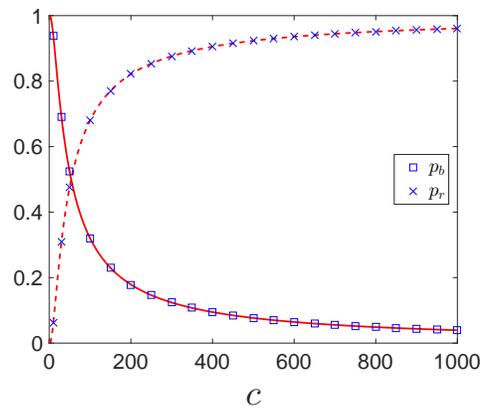}
}
\caption{
The probabilities 
$p_b$ 
and 
$p_r$
that a RW on an ER network will terminate via backtracking or
by retracing its path, respectively,
as a function of the mean degree, $c$.
The theoretical results, obtained from Eqs.
(\ref{eq:p_b2})
and
(\ref{eq:p_r2})
are found to be in excellent agreement with 
the results of numerical simulations 
(symbols).
}
\label{fig:7}
\end{figure}

In Fig.
\ref{fig:7}
we present the probability $p_b$ that the RW will terminate due to backtracking
and the probability $p_r$ that it will terminate due to retracing as a function
of the mean degree $c$ for an ER network of size $N=1000$.
As expected, $p_b$ is a decreasing function of $c$ while $p_r$
is an increasing function.
The two curves intersect at 
$c=c^{\ast}$,
where
$p_b(c^{\ast}) = p_r(c^{\ast}) = 1/2$.
To evaluate $c^{\ast}$ we use Eq.
(\ref{eq:p_b2}).
Since this crossover is expected to take place at a large 
value of $c$ we can plug in the expression for 
$\ell_{mean}$ from
Eq.
(\ref{eq:ell_mean5})
and obtain
$p_b \simeq  ( \pi N )^{1/2} /{c}$.
Therefore, the crossover takes place at
$c^{\ast} \simeq \left(  \pi N \right)^{1/2}$.
For the network size presented here, of $N=1000$, the 
crossover point is predicted to be at $c^{\ast} = 57$,
in agreement with the numerical results. 

\begin{figure}
\centerline{
\includegraphics[width=12cm]{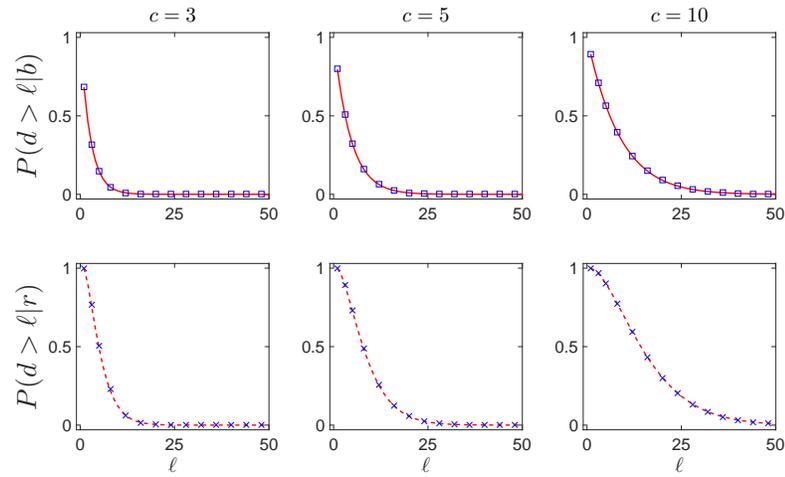}
}
\caption{
The conditional tail distributions 
$P(d>\ell | b)$
and
$P(d>\ell | r)$
of first hitting times 
vs. $\ell$
for RWs on an ER network, 
for paths terminated by backtracking
(top row) or by retracing (bottom row),
respectively.
The results are shown for $N=1000$
and $c=3$, $5$ and $10$.
The theoretical results for 
$P(d>\ell | b)$ 
are obtained from Eq.
(\ref{eq:p_lbgt}),
while the theoretical results for
$P(d>\ell | r)$ 
are obtained from Eq.
(\ref{eq:p_lrgt}).
In both cases, the theoretical results (solid lines) are
found to be in excellent agreement with the numerical simulations
(symbols).
}
\label{fig:8}
\end{figure}

In Fig.
\ref{fig:8}
we present the probabilities
$P(d>\ell | b)$
and
$P(d>\ell | r)$
that a RW will have a path of length larger than $\ell$ 
given that it terminates due to backtracking or retracing,
respectively. The results are presented for $N=1000$ and
$c=3$, $5$ and $10$.
The analytical results (solid lines) are found to be in excellent agreement
with the numerical simulations (circles).
In both cases, the paths tend to become longer as $c$ is increasd.
However, for each value of $c$, the paths which terminate by retracing
are typically longer than the paths which terminate by backtracking.

\begin{figure}
\centerline{
\includegraphics[width=12cm]{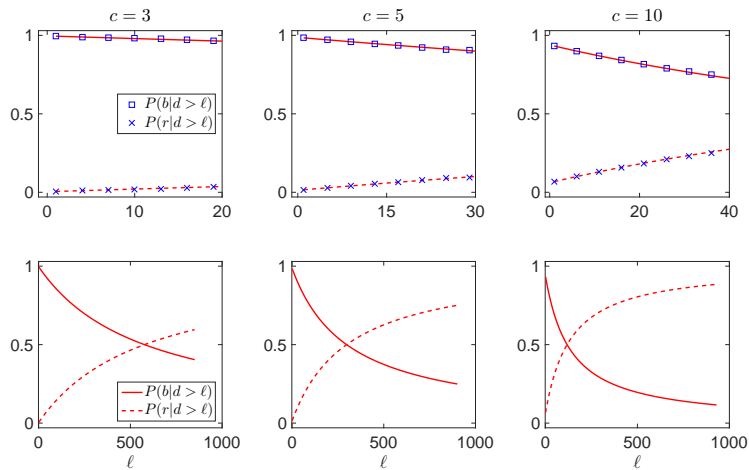}
}
\caption{
The conditional probabilities
$P(b | d>\ell)$
and
$P(r | d>\ell)$
that a RW path will terminate by backtracking or by retracing,
respectively, given that its length is larger than $\ell$,
are presented as a function of $\ell$.
Results are shown for an ER networks of size $N=1000$ and 
$c=3$, $5$ and $10$.
The theoretical results for
$P(b | d>\ell)$ (solid lines)
are obtained from Eq.
(\ref{eq:b_ell3})
while the theoretical results for
$P(r | d>\ell)$ (dashed lines)
are obtained from Eq.
(\ref{eq:r_ell}).
In the top row these results are compared to the results of numerical
simulations (symbols) finding excellent agreement. This comparison
is done for the range of path lengths which actually appear in the
numerical simulations and for which good statistics can be obtained.
Longer RW paths which extend beyond this range become extremely
rare, so it is difficult to obtain sufficient numerical data.
However, in the bottom row we show the theoretical results
for the entire range of path lengths. 
It is found that 
$P(b | d>\ell)$
is a monotonically decreasing function of $\ell$
while
$P(r | d>\ell)$ is monotonically increasing.
The two curves intersect each other at a value $\ell$
where both probabilities are equal to $1/2$.
}
\label{fig:9}
\end{figure}

In Fig. 
\ref{fig:9}
we present the probabilities
$P(b | d>\ell)$
and
$P(r | d>\ell)$
that a RW will terminate due to backtracking or retracing,
respectively.
Results are shown for ER networks of size $N=1000$ and 
$c=3$, $5$ and $10$.
The theoretical results for
$P(b | d>\ell)$ (solid lines)
are obtained from Eq.
(\ref{eq:b_ell})
while the theoretical results for
$P(r | d>\ell)$ (dashed lines)
are obtained from Eq.
(\ref{eq:r_ell}).
As expected,
it is found that 
$P(b | d>\ell)$
is a monotonically decreasing function of $\ell$
while
$P(r | d>\ell)$ is monotonically increasing.
In the top row these results are compared to the results of numerical
simulations (symbols) finding excellent agreement. 
This comparison
is done for the range of path lengths which actually appear in the
numerical simulations.
Longer RW paths which extend beyond this range become extremely
rare, so it is difficult to obtain sufficient numerical data.
However, in the bottom row we show the theoretical results
for the entire range of path lengths. 
In fact, such long paths can be sampled using the pruned 
enriched Rosenbluth method, which was successfully used in the 
context of SAWs in polymer physics 
\cite{Grassberger1997}. 
In this method one samples long non-overlapping paths, 
keeping track of their weights, to obtain an unbiased sampling 
in the ensemble of all paths.

\section{Summary and Discussion}

We have presented analytical results for   
the distribution of first hitting times of random walkers on 
ER networks.
Starting from a random initial node,
these walkers hop randomly between adjacent nodes 
until they hit a node which they already visited before.
At this point, the path is terminated.
The number of steps taken from the initial node
up to the termination of the path is called the first
hitting time.
Using recursion equations, 
we obtained 
analytical results for the distribution of first hitting times, 
$P(d=\ell)$.
One can distinguish between two termination scenarios,
referred to as backtracking and retracing.
We have performed a detailed analysis of the probabilities,
$p_b$ and $p_r$,
that the termination will take place via the backtracking or via
the retracing mechanism, respectively.
We obtained analytical expressions for these probabilities in terms
of the network size, $N$ and the mean degree, $c$.
We also obtained analytical expressions for the conditional 
distributions of the path lengths,
$P(d=\ell | b)$ 
and
$P(d=\ell | r)$
for the paths which terminate by
backtracking and by retracing,
respectively.
Finally, we obtained analytical expressions for 
the conditional probabilities
$P(b | d=\ell)$
and
$P(r | d=\ell)$
that a path which terminates after $\ell$ steps
is terminated by backtracking or by retracing, respectively.
It was found that the two termination mechanisms exhibit
very different behavior.
The backtracking probability sets in starting from the 
second step and is constant throughout the path.
As a result, this mechanism alone whould produce a 
geometric distribution of path lengths.
The retracing mechanisms sets in starting from the 
third step and its rate increases linearly in time.
The balance between the two termination mechanisms 
depends on the mean degree
of the network.
In the limit of sparse networks, 
the backtracking mechanism is dominant and most
paths are terminated long before the retracing mechanism becomes relevant.
In the case of dense networks, the backtracking probability is low and most
paths are terminated by the retracing mechanism.

In Table 1 we summarize the main results of the paper, providing links
to the corresponding equations, for three different levels of precision.
The results which are given by closed form expressions appear in boldface fonts.
The first column includes exact results. Most of these results are expressed in terms of 
sums and products of the conditional probabilities, with no closed form expression. 
The second column includes the results
obtained by replacing the sums by integrals.
These results are of high accuracy since the relative errors invloved in the
replacement of the sums by integrals are small and scale like
$\ell/N$.
The third column includes approximate results, which are given by
closed form expressions. 
These results are obtained from a series expansion truncated above
the second order in $\ell/N$, and the errors involved in this approximation
scale like $\ell^3/N^2$. 
The conditional probabilities of the two termination
scenarios are expressed in terms of $P(d>\ell)$ and $\ell_{\rm mean}$.
Therefore, the level of precision of these conditional probabilities depends
on the precision used for these input functions, as indicated by the superscripts. 

Under conditions in which backtracking and retracing steps are not 
allowed the RW becomes an SAW. 
It is terminated via the stalemate
scenario, when all the nodes surrounding the current node have already been
visited.  The resulting path length is referred to as the last hitting time.
In Ref. 
\cite{Tishby2016}
it was shown that for
large networks 
($N \gg 1$)
the tail distribution
of the last hitting times
denoted by
$P_{\rm L}(d>\ell)$,
follows a Gompertz distribution 
\cite{Gompertz1825,Johnson1995,Shklovskii2005}
of the form

\begin{equation}
P_{\rm L} \left(d>\ell\right)
\simeq
\exp\left[-\frac{N}{c}e^{-c}
\left(e^{\frac{c}{N}\ell}-1\right)\right],
\label{eq:tail2}
\end{equation}

\noindent
and the corresponding probability density is given by

\begin{equation}
P_{\rm L} (d=\ell)
\simeq
\exp\left[-\frac{N}{c}e^{-c}
\left(e^{\frac{c}{N}\ell}-1\right)
-\left(1-\frac{\ell+1}{N}\right)c
\right].
\label{eq:pdf2}
\end{equation}

\begin{figure}
\centerline{
\includegraphics[width=7cm]{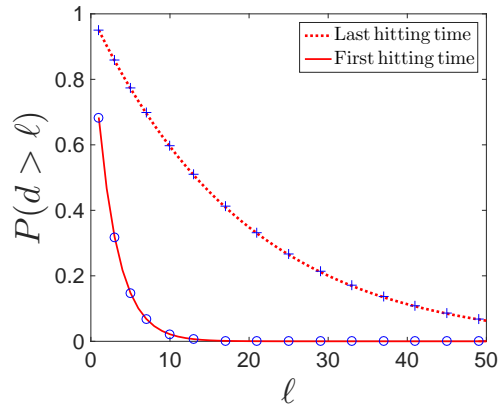}
}
\caption{
Theoretical results for the tail distribution
$P(d > \ell)$ 
of the first hitting times of the RW model
(solid line)
and the tail distribution
$P_{\rm L}(d>\ell)$
of the last hitting times of the SAW model
(dashed line),
on ER networks
of size $N=1000$ 
and 
$c=3$.
Both results are in excellent agreement
with numerical simulations (symbols).
As expected, the last hitting times are much
longer than the first hitting times.
}
\label{fig:10}
\end{figure}

\noindent
In Fig.
\ref{fig:10}
we present the tail distribution
$P(d>\ell)$ 
of first hitting times
(solid line)
and the tail distribution
$P_{\rm L}(d>\ell)$
of last hitting times
(solid line)
for an ER network of size
$N=1000$ and mean degree
of $c=3$.
As expected, there is a large gap between the first hitting time and the 
last hitting time. This gap increases as the network becomes denser.

Beyond the specific problem of first hitting times of RW on networks,
the analysis presented here provides useful insight into the general
context of the distribution of life expectancies of humans, animals and machines
\cite{Finkelstein2008,Gavrilov2001}.
It illustrates the combination of two lethal hazards, where one hits at a fixed,
age-independent rate, while the other increases linearly with age.
The first hazard may be considered as an external cause such as an accident
while the second hazard involves some aging related 
degradation which results
in an increasing failure rate.

\newpage

\begin{table}
\caption{Summary of the main results and their level of precision}
\begin{center}
\begin{tabular}{| c  | l | l | l |}
\hline \hline
{\bf Property}  & {\bf Exact} & {\bf Accurate}  & {\bf Approximate}  \\ 
     \hline \hline 
$p_t(k)$     &  {\bf Eq. (\ref{eq:p_t(k)})$^{\ast}$ }     & -- &  -- \\ \hline
$c(t)$    &  {\bf Eq. (\ref{eq:coft})  }   &  --   &  --  \\ \hline \hline
$P_b(d>\ell)$   & {\bf Eq. (\ref{eq:P_b1})  }   &  --   & -- \\ \hline
$P_r(d>\ell)$   &  Eq. (\ref{eq:P_rsum})     &  {\bf Eq. (\ref{eq:P_r4}) }  & {\bf Eq. (\ref{eq:tail_simp2}) } \\ \hline
$P(d>\ell)$   &  Eq. (\ref{eq:cond2})$^{1}$     &  {\bf Eq. (\ref{eq:cond2})$^{2}$}  & {\bf Eq. (\ref{eq:tail_simp3})}   \\ \hline \hline
$\ell_{\rm mean}$ &  Eq. (\ref{eq:tailsum2})  & Eq. (\ref{eq:ell_mean1})$^{2}$ & {\bf Eq. (\ref{eq:ell_mean5}) } \\ \hline
$\langle \ell^2 \rangle$ &   --    &  --  & {\bf Eq. (\ref{eq:ell_sqr})  } \\ \hline
$\sigma_{\ell}^2(c)$ &  --     & --    &  {\bf Eq. (\ref{eq:sigmaell}) }\\ \hline \hline
$p_b$ &  Eq. (\ref{eq:p_b2})$^{a}$     &  Eq. (\ref{eq:p_b2})$^{b}$    & {\bf Eq. (\ref{eq:p_b2})$^{c}$ } \\ \hline
$p_r$ &   Eq. (\ref{eq:p_r2})$^{a}$     & Eq. (\ref{eq:p_r2})$^{b}$    & {\bf Eq. (\ref{eq:p_r2})$^{c}$ } \\ \hline
$P(d>\ell | b)$  &   Eq. (\ref{eq:p_lbgt})$^{1,a}$   &   Eq. (\ref{eq:p_lbgt})$^{2,b}$    & {\bf  Eq. (\ref{eq:p_lbgt})$^{3,c}$  }     \\ \hline
$P(d>\ell | r)$         &  Eq. (\ref{eq:p_lrgt})$^{1,a}$     &   Eq. (\ref{eq:p_lrgt})$^{2,b}$      & {\bf Eq. (\ref{eq:p_lrgt})$^{3,c}$ } \\ \hline
$P(b | d>\ell)$  & Eq. (\ref{eq:b_ell3})$^{1}$     &     Eq. (\ref{eq:b_ell3})$^{2}$   &  {\bf  Eq. (\ref{eq:b_ell3})$^{3}$   }     \\ \hline
$P(r | d>\ell)$          &    Eq. (\ref{eq:r_ell})$^{1}$    &  Eq. (\ref{eq:r_ell})$^{2}$       &  {\bf   Eq. (\ref{eq:r_ell})$^{3}$ } \\ \hline \hline
\end{tabular}
\end{center}
$^{\ast}$boldface fonts: closed-form expressions;
\\
$^{1}$where $P_r(d>\ell)$ is given by Eq.  (\ref{eq:P_rsum}); 
\\
$^{2}$where $P_r(d>\ell)$ is given by Eq.  (\ref{eq:P_r4}); 
\\
$^{3}$where $P_r(d>\ell)$ is given by Eq.  (\ref{eq:tail_simp2}); 
\\
$^{a}$where $\ell_{\rm mean}$ by Eq. (\ref{eq:tailsum2});
\\
$^{b}$where $\ell_{\rm mean}$ by Eq. (\ref{eq:ell_mean1});
\\
$^{c}$where $\ell_{\rm mean}$ by Eq. (\ref{eq:ell_mean5});
\label{table}
\end{table}

\newpage

\end{document}